\documentclass[aps,onecolumn,showpacs,nofootinbib,showkeys]{revtex4-2}
\usepackage[margin=3.15cm]{geometry}

\RequirePackage[T1]{fontenc}

\usepackage{epsfig,graphicx,amsmath,amssymb,bm}
\RequirePackage{mathptmx}      
\usepackage{mathrsfs}
\usepackage{amssymb}
\RequirePackage{color}
\usepackage{changes}
\usepackage{slashed}
\usepackage{multirow}
\usepackage{scalerel}
\usepackage{tikz-feynman}
\usepackage{textcomp}
\usepackage{subcaption}
\usepackage[english]{babel}
\usepackage{float}
\RequirePackage{hyperref}
\hypersetup{
    linktocpage,
    colorlinks,
    citecolor=blue,
    filecolor=black,
    linkcolor=blue,
    urlcolor=blue,
}
\usepackage{nccmath}

\begin{document}

\title{Chiral anomaly in the $\eta^{(\prime)}\to\pi^+\pi^-\gamma$ and $\eta^{(\prime)}\to\pi^+\pi^-l^+l^-$ decays} 

\author{M.~K. Volkov$^{1}$}\email{volkov@theor.jinr.ru}
\author{A.~A. Osipov$^{1}$}\email{aaosipov@theor.jinr.ru}
\author{K.~Nurlan$^{1,2}$}\email{nurlan@theor.jinr.ru}
\author{A.~A. Pivovarov$^{1}$}\email{pivovarov@theor.jinr.ru}

\affiliation{$^1$ Bogoliubov Laboratory of Theoretical Physics, JINR, 
                 141980 Dubna, Moscow region, Russia \\
                $^2$ The Institute of Nuclear Physics, Almaty, 050032, Kazakhstan}    



\begin{abstract}
We report the presence of a flavor-violating correction to the $VAAA$-type anomaly, $\delta^{(\prime)}$, induced by the surface terms of the anomalous quark triangle diagram, previously found in the $\eta,\eta'\to\pi^+\pi^-\gamma$ decay amplitudes, and investigate its impact on the corresponding semileptonic decay modes of $\eta$ and $\eta'$. The magnitude of $\delta^{(\prime)}$ can be set from the experimental data on the $\eta^{(\prime)}\to \pi^+\pi^-\gamma$ decay width. We then estimate another low-energy constant, the slope parameter $\alpha^{(\prime)}$. The impact of different schemes for describing $\eta$-$\eta^\prime$ mixing on the value of $\delta^{(\prime)}$ and $\alpha^{(\prime)}$ is discussed. The predictions are shown to be in complete agreement with the available experimental data.
\end{abstract}



\maketitle
\large
\section{Introduction}
In a recent paper \cite{Osipov:2020vad}, we analyzed the anomalous $\eta^{(\prime)}\to\pi^+\pi^-\gamma$ decays using the approach based on the bosonized version of the Nambu-Jona-Lasinio (NJL) model \cite{Nambu:61a,Nambu:61b}, which explicitly includes scalar, pseudoscalar, vector and axial-vector mesons \cite{Eguchi:74,Kikkawa:76,Ebert:1982pk, Volkov:1984kq, Volkov:1986zb,Ebert:1985kz,Vogl:1991qt,Klevansky:1992qe,Volkov:1993jw,Bijnens:93, Hatsuda:1994pi, Ebert:1994mf, Volkov:2005kw,Osipov:2006,Osipov:2007}. Here we extend the calculation to the related decays $\eta^{(\prime)}\to\pi^+\pi^-l^+l^-$. While the $\eta^{(\prime)}\to\pi^+\pi^-\gamma$ modes allow us to deepen our understanding of the low-energy structure of the hadronic part of the amplitude far from the chiral limit, the analysis of dilepton modes additionally enable us to probe in more detail the hadronic form factor $\mathcal F_{\eta^{(\prime)}}(q^2,p_+,p_-)$, defined as 
\begin{equation}
\langle \pi^+(p_+)\pi^-(p_-) | J_\mu | \eta^{(\prime)}(p_{\eta^{(\prime)}}) \rangle =e_{\mu\nu\alpha\beta}q^\nu p_+^\alpha p_-^\beta \mathcal F_{\eta^{(\prime)}} (q^2,p_+,p_-),
\end{equation}
under conditions of non-zero photon virtuality $q^2 > 0$. Here $p_\pm$, $p_{\eta^{(\prime)}}$ are the momenta of the charged pions and the $\eta^{(\prime)}$ meson, and $J_\mu=e(2\bar u\gamma_\mu u-\bar d\gamma_\mu d-\bar s\gamma_\mu s)/3$ is the electromagnetic current of quarks, where $e$ is the electric charge. 

The $\eta^{(\prime)}\to\pi^+\pi^-l^+l^-$ decays are widely studied in the literature both within a chiral unitary approach which combines the chiral perturbation theory (ChPT) Lagrangian with a coupled-channels Bethe-Salpeter equation \cite{Borasoy:2007} and within the framework of the combined use of ChPT and dispersion analysis methods \cite{Kubis:22}. Alternatively, model approaches based on effective chiral Lagrangians which incorporate vector and axial-vector mesons can be used \cite{Meissner:88}. These include the theoretical description of spin-1 mesons as massive Yang-Mills (MYM) bosons \cite{Schechter:84,Schechter:85}, and the approach in which vector states are dynamical gauge bosons of hidden local symmetry (HLS) \cite{Bando:85,Bando:85b}. In particular, the calculation of the decay width $\eta\to\pi^+\pi^- e^+ e^-$ and differential distributions in terms of invariant masses $\pi^+\pi^-$ and $e^+ e^-$, performed within the framework of the HLS approach, is presented in \cite{Picciotto:93}. Our approach is closest to the latter ones. The main difference is that we include the contribution generated by the surface term to the $\eta^{(\prime)}\to\pi^+\pi^-\gamma$ and $\eta^{(\prime)}\to\pi^+\pi^-l^+l^-$ amplitudes, which is absent in \cite{Picciotto:93}. The appearance of the surface term is related to the quark nature of meson states. In the NJL model, which directly uses the language of quarks and mesons, this contribution is associated with the triangle quark diagram of type AAA (A -- axial-vector current). Due to the formal linear divergence of the loop integral, a shift in the integration momentum changes its magnitude \cite{Jackiw:00,Jackiw:72}. The resulting ambiguity, as shown in \cite{Osipov:18a,Osipov:18b,Osipov:20}, can be used to solve the well-known problem associated with the violation of Ward's chiral identities due to pseudoscalar -- axial-vector (PA) mixing in the anomalous part of the effective Lagrangian. This also leads to an additional contribution to the $VAAA$ box anomaly (V -- vector current), the role of which is the main subject of this paper. 

A key feature of the NJL approach is that the vertices of the effective meson Lagrangian are obtained by integrating out the quark degrees of freedom of the corresponding effective action and then expanding the modulus of the resulting quark determinant in the low-energy (long wavelength) regime using the heat kernel technique (i.e. an expansion in powers of derivatives of the meson fields) \cite{Ebert:1985kz,Bijnens:93}. This method allows one to isolate local contributions from quark one-loop diagrams without destroying the chiral structure of non-anomalous meson vertices. The divergent parts of the latter are regularized by introducing a cutoff $\Lambda$ at energies of the order of the spontaneous chiral symmetry breaking scale $\Lambda\simeq 4\pi f_\pi \sim 1\,\mbox{GeV}$ \cite{Manohar:84}. This circumvents the fundamental problem of imaginary parts of quark loop diagrams, the presence of which would indicate quark deconfinement. The long wavelength expansion of one-loop quark diagrams is the main approximations of the approach we use here. It is in this approximation of the NJL model that one can obtain effective meson Lagrangians for various well-known approaches, such as MYM and HLS. Moreover, it is in this approximation that the NJL model predictions agree well with the results of ChPT \cite{Arriola:91, Bijnens:93}.

The presence of chiral anomalies leads to an imaginary part of the NJL based effective action which is of the type considered by Wess-Zumino and Witten (WZW) \cite{WZ:71,Witten:83}. It follows that the general structures of the MYM and HLS approaches, except for the numerical values of the parameters, are equivalent not only in the non-anomalous but also in the anomalous sectors and can be obtained from the NJL model, as demonstrated in \cite{Wakamatsu:89}. 
Therefore, at first glance, it might seem that using the NJL model  would add little to the results, already known from MYM and HLS calculations. This impression is erroneous. Indeed, the results on $\eta\to\pi^+\pi^- \gamma$ of \cite{Picciotto:92} are based on rigorous $SU(3)\times SU(3)$ low-energy theorems and take into account the effects of explicit chiral symmetry breaking only in that part of the hadron form factor $\mathcal F_{\eta}(0,p_+,p_-)$ that owes its origin to $\eta$-$\eta^\prime$ mixing. New experimental data however  indicate that this is clearly insufficient: A more thorough study of the $\eta^{(\prime)}\to\pi^+\pi^-\gamma$ decay amplitude has demonstrated the importance of taking into account the effects of explicit symmetry breaking not only when describing $\eta$-$\eta^\prime$ mixing \cite{Benayoun:03}, but also directly in the perturbative (in the ChPT sense) part of the hadron form factor \cite{Osipov:2020vad,Benayoun:10}. This is the question we will address in this work.

Let us dwell on this in more detail. Since the pion pair in the $\eta^{(\prime)}\to\pi^+\pi^-\gamma$ decays is produced predominantly in the $p$-wave $1^{--}$ state, it can be assumed (and this assumption is confirmed experimentally) that the momentum-dependent hadronic part of the form factor can be represented as a product of the polynomial $P_{\eta^{(\prime)}}(s_{\pi\pi})$ and the pion vector form factor $F_V(s_{\pi\pi})$ with the standard normalization $F_V(0)=1$ \cite{Meissner:12}
\begin{equation}
\label{Gampl}
\mathcal F_{\eta^{(\prime)}}(0,p_+,p_-)\to \mathcal F_{\eta^{(\prime)}}(0,s_{\pi\pi}) = P_{\eta^{(\prime)}}(s_{\pi\pi}) F_V(s_{\pi\pi}).
\end{equation} 

The polynomial $P_{\eta^{(\prime)}}(s_{\pi\pi})$, where $s_{\pi\pi}=(p_++p_-)^2$, is reaction specific and for small values of $s_{\pi\pi}$ is expected to be perturbative in the sense of ChPT.  A comparison with experimental data obtained by the WASA-at-COSY \cite{WASA:2012}, KLOE \cite{KLOE:2013} and CRYSTAL BARREL \cite{Abele:97} Collaborations shows that, within experimental accuracy, the polynomial $P_{\eta^{(\prime)}}(s_{\pi\pi})$ can be considered linear in the invariant mass $s_{\pi\pi}$ of the pion pair,   
\begin{equation}
\label{Pol}
P_{\eta^{(\prime)}}(s_{\pi\pi})=A_{\eta^{(\prime)}} (1+\delta^{(\prime)}) (1+\alpha^{(\prime)} s_{\pi\pi}),
\end{equation}
with 
\begin{eqnarray}
\label{WASA}
\alpha &=&(1.89\pm 0.25_{stat}\pm 0.59_{sys}\pm 0.02_{theo})\,\mbox{GeV}^{-2} \quad\mbox{\cite{WASA:2012}}, \\
\label{KLOE}
\alpha &=&(1.32\pm 0.08_{stat} \left.^{+0.10}_{-0.09}\right._{syst} \pm 0.02_{theo})\,\mbox{GeV}^{-2}  \quad \ \,\mbox{\cite{KLOE:2013}}, \quad \\
\label{C-Bc}
\alpha^\prime &=&(1.80\pm 0.49\pm 0.04)\,\mbox{GeV}^{-2} \qquad\qquad\qquad  \mbox{\cite{Meissner:12} and \cite{Abele:97}}.
\end{eqnarray}
The slope parameter $\alpha'$ was obtained in \cite{Meissner:12} by fitting the spectral data of the CRYSTAL BARREL collaboration \cite{Abele:97}, where the first error is introduced by the data on the decay $\eta'\to\pi^+\pi^-\gamma$, and the second by the data on the vector form factor of the pion.

It follows from (\ref{Gampl}) that the hadron form factor $\mathcal F_{\eta^{(\prime)}}(0,s_{\pi\pi})$ satisfies the low energy theorem.
\begin{equation}
\label{N}
\mathcal F_{\eta^{(\prime)}}(0,0)=A_{\eta^{(\prime)}}(1+\delta^{(\prime)}).
\end{equation}
The factor $A_{\eta^{(\prime)}}$ in the chiral limit is completely determined by the WZW Lagrangian \cite{WZ:71, Witten:83}. Taking into account singlet-octet mixing, it has the form
\begin{equation}
\label{A}
A_{\eta^{(\prime)}}=\frac{eN_c}{12\pi^2 f_\pi^3} c_{\eta^{(\prime)}}.
\end{equation}
Here $N_c$ is the number of quark colors, $f_\pi=92.2\, \mbox{MeV}$ is the pion decay constant, and the constants $c_\eta$, $c_{\eta'}$ depend on the $\eta$-$\eta'$ mixing scheme. For exact $U(3)$ symmetry $c_{\eta}=1/\sqrt 3$, $c_{\eta'}=\sqrt{2/3}$, as required by the chiral anomaly. 

Thus, in addition to the $\eta$-$\eta'$ mixing parameters $c_{\eta^{(\prime)}}$, the hadronic form factor contains at least two more low-energy parameters, $\delta^{(\prime)}$ and $\alpha^{(\prime)}$. These parameters accumulate important information related to the mechanism of the decay process. According to the Adler-Bardeen nonrenormalizability theorem, the parameter $\delta^{(\prime)}$ in (\ref{N}) must vanish in the chiral limit, otherwise the consequence of the anomalous Ward identities, Eq.(\ref{A}), will be distorted. Therefore, it is this parameter that absorbs the chiral corrections associated with the explicit violation of flavor $SU(3)$ symmetry, which are related to the box VAAA-anomaly. As will be demonstrated below, the parameter $\delta^{(\prime)}$ is non-zero, vanishes in the chiral limit, and has a noticeable effect on the slope parameter $\alpha^{(\prime)}$. Thus, one of the goals of this work is to demonstrate the importance and possible experimental manifestations of this contribution in $\eta^{(\prime)}\to\pi^+\pi^-\gamma$ and $\eta^{(\prime)}\to\pi^+\pi^-l^+l^-$ decays.

The paper is organized as follows. In Sect.\,II, we discuss in detail the origin of the parameter $\delta^{(\prime)}$ in the NJL model. We briefly outline the most important details necessary for understanding the main steps of our calculations. These include the Lagrangian of the model, the schemes used to describe the $\eta$-$\eta^\prime$ mixing, aspects of deriving the $\eta^{(\prime)}\to\pi^+\pi^-\gamma$ amplitudes. In Sect.\,III, we discuss the limitations of the NJL model approach and, in particular, justify subsequent steps that take us beyond NJL calculations and thus allow us to move forward in our phenomenological analysis. Here we find the relation between $\delta^{(\prime)}$ and slope parameter $\alpha^{(\prime)}$, and dwell on the clarification of the physical content of the contributions constituting $\alpha^{(\prime)}$. The $\eta^{(\prime)}\to\pi^+\pi^-l^+l^-$ decay amplitudes, numerical estimates of the corresponding decay widths and distributions in terms of the $\pi^+\pi^-$ and $e^+e^-$ invariant masses are presented in Sect.\,IV. In Sect.\,V, we discuss the obtained results and highlight the most important ones.


\section{Origin and role of the parameter $\delta^{(\prime )}$ in the NJL model}

Let us clarify how the low-energy parameter $\delta^{(\prime)}$ appears in the structural part of the amplitude $\eta/\eta^\prime\to\pi^+\pi^-\gamma$. Although this was partially described in our previous paper \cite{Osipov:2020vad}, here we present additional arguments based on the NJL model and show that isospin symmetry breaking already at the level of  tree diagrams leads to a non-zero value of $\delta^{(\prime)}$. We hope that this will in some sense complement the picture presented in \cite{Meissner:12}, where the matching of the $\eta/\eta^\prime\to\pi^+\pi^-\gamma$ decay amplitude  to one-loop ChPT result showed that $\delta^{(\prime)}$ can be associated with the contribution of one-loop meson diagrams. We would like to emphasize that all our numerical estimates made in the following sections will be still based on the results of the work \cite{Osipov:2020vad}. However, the analysis presented below, in addition to demonstrating the origin of the parameter $\delta^{(\prime)}$, may in the future serve as a basis for conducting consistent calculations  beyond the tree approximation within the NJL model.

\subsection{Lagrangian}
Our starting point is the chiral Lagrangian, which is an extended  quark version of the NJL model with $U(3)_L\times U(3)_R$ symmetry \cite{Volkov:1984kq,Volkov:1986zb, Ebert:1985kz}
\begin{equation}
\mathcal L= \bar q [i\gamma^\mu (\partial_\mu -ieQ\mathcal A_\mu ) -m]q +L_{int} +L_{em}, 
\end{equation}
where $Q=\mbox{diag}(2,-1,-1)/3$ is the diagonal matrix of $SU(3)$ quark charges, $q$ and $\mathcal A_\mu$ are the quark and electromagnetic fields, $L_{em}$ is the Lagrangian density of the free electromagnetic field, $\gamma^\mu$ are the Dirac matrices, the diagonal matrix $m = \mbox{diag}(m_u , m_d , m_s)$ contains the current masses of the $u$, $d$, and $s$ quarks. The Lagrange density $L_{int} = L^{(0)} + L^{(1)}$, where the sum includes $U(3)_L \times U(3)_R$ chiral-symmetric combinations describing four-quark interactions with spin zero and one
\begin{eqnarray}
L^{(0)}&=&\frac{G_S}{2}\left[(\bar q\lambda_a q)^2 + (\bar q i\gamma_5 \lambda_a q)^2 \right], \\
L^{(1)}&=&-\frac{G_V}{2}\left[(\bar q \gamma_\mu \lambda_a q)^2 + (\bar q \gamma_\mu \gamma_5 \lambda_a q)^2 \right].
\end{eqnarray}
Here $G_S$, $G_V$ are universal quark coupling constants with a dimension $[G_{S,V}]=\mbox{mass}^{-2}$; $\lambda_a$ are the generators of the flavor $U(3)$ group, normalized as $\langle \lambda_a\lambda_b\rangle =2\delta_{ab}$ with $a,b=0,1,\ldots 8$ (where $\langle\ldots\rangle \equiv \mbox{tr}_F(\ldots )$, trace over flavor indices),  and $\lambda_0 =\sqrt{2/3}\, \textbf{1}$ with $\textbf{1}$ being a unit $3\times 3$ matrix. Following \cite{Kikkawa:76}, we can introduce auxiliary fields $\sigma_a=G_S(\bar q\lambda_a q)$, $\phi_a=G_S(\bar q i\gamma_5 \lambda_a q)$, $V_{\mu a}=G_V(\bar q \gamma_\mu\lambda_a q)$, $A_{\mu a}=G_V(\bar q \gamma_\mu \gamma_5 \lambda_a q)$ and thereby move to an equivalent (in the sense of the functional integral) theory with a Lagrangian density
\begin{equation}
\label{Lqm}
\mathcal L' =\bar q Dq-\frac{1}{4G_S}\langle \sigma^2+\phi^2 \rangle +\frac{1}{4G_V}\langle V_\mu^2+A_\mu^2\rangle +L_{em}
\end{equation} 
with $D$ being a Dirac operator in the presence of bosonic fields $\sigma=\sigma_a\lambda_a$, $\phi=\phi_a\lambda_a$, $V_\mu =V_{\mu a} \lambda_a$, $A_\mu =A_{\mu a} \lambda_a$, namely
\begin{equation}
\label{D}
D=i\gamma^\mu d_\mu -m+\sigma +i\gamma_5\phi, \quad d_\mu=\partial_\mu-ieQ\mathcal{A_\mu} -iV_\mu-i\gamma_5 A_\mu.
\end{equation}

It is known \cite{Nambu:61a} that if the constant $G_S$ exceeds a certain critical value $G_S^{crit}$, an important phenomenon occurs, caused by a vacuum rearrangement due to spontaneous chiral symmetry breaking. As a result, a gap appears in the fermion spectrum. Formally, this is achieved by shifting the scalar field $\sigma\to\sigma -M+m$ in (\ref{Lqm}), where $M =\mbox{diag}(M_u, M_d, M_s)$ is the mass matrix of the constituent quarks. The latter are determined from the gap equation 
\begin{equation}
\label{gap}
M_i\left[1-8 G_S I_1(M_i)\right]=m_i, 
\end{equation}
where
\begin{equation}
\quad I_1(M_i)=\frac{i N_c}{(2\pi)^4}\int_\Lambda \frac{d^4p}{p^2-M_i^2}=\frac{N_c}{(4\pi)^2}\left[
\Lambda^2-M_i^2 \ln\left(1+\frac{\Lambda^2}{M^2_i}\right)\right].
\end{equation}
Here, after moving to Euclidean space, we use the covariant cutoff at scale $\Lambda$ to regularize this quadratically divergent integral. Clearly, in the strong coupling regime
\begin{equation}
G_S\Lambda^2>\frac{2\pi^2}{N_c}=6.58
\end{equation}
each of the three equations (\ref{gap}) has a nontrivial solution describing the gap in the fermion spectrum. This solution indicates that the ground state becomes superconducting with a nonzero quark condensate $\langle 0|\bar q\lambda_i q|0\rangle =-(M_i-m_i)/(2G_S)$.

Small fluctuations of meson fields in the superconducting ground state are described by the effective Lagrangian obtained from (\ref{Lqm}) by shifting $\sigma\to\sigma -M+m$ and then integrating out quark fields, which, as is well known, leads to the quark determinant of the Dirac operator (\ref{D}). The result is a nonlocal functional of collective meson fields
\begin{eqnarray}
\label{L2}
\mathcal L''&=&-i\mbox{Tr}\ln\left[1+(i\gamma^\mu\partial_\mu -M)^{-1}\left(\sigma+i\gamma_5\phi +\gamma^\mu(eQ\mathcal{A_\mu} +V_\mu+\gamma_5 A_\mu) \right)\right]_\Lambda \nonumber \\
&-&\frac{1}{4G_S}\langle \left(\sigma-M+m\right)^2+\phi^2 \rangle +\frac{1}{4G_V}\langle V_\mu^2+A_\mu^2\rangle +L_{em},
\end{eqnarray}
where $\mbox{Tr}$ means the trace operation with respect to the space-time points, $\gamma_\mu$-matrices, color and flavor indices. A detailed technique for such calculations is developed in \cite{Osipov:96}. It has proven itself in describing $\pi\pi$ scattering; however, for problems of $\eta$ and $\eta'$ physics, it leads to the appearance of imaginary parts associated with quark deconfinement. To circumvent the deconfinement problem, one usually resorts to a low-energy (long-wavelength) proper time expansion of the fermion determinant \cite{Ball:89}. There are a number of possibilities here, from the classical Schwinger method used to extract the infinities contained in the real part of the one-loop action \cite{Schwinger:51} to its extension to the case of unequal quark masses in the loop \cite{Osipov:21plb,Osipov:21prd}.
In the following we choose a less rigorous but intuitively sound approach, used in \cite{Kikkawa:76,Volkov:1986zb, Ebert:1985kz}. 

To transition to physical states, we need to consider the part of the Lagrangian (\ref{L2})  that is quadratic in the fields, isolate the first terms in the derivative expansion (they diverge as $\Lambda\to\infty$), and reduce them to canonical form. The latter leads to a redefinition of the collective fields
\begin{equation}
\label{gV}
\sigma =g_{\sigma}\circ \sigma',\quad  \phi=g_\sigma\circ \phi', \quad V_\mu=\frac{g_V}{2}\circ V_\mu', \quad A_\mu=\frac{g_V}{2}\circ A_\mu' .
\end{equation}
For convenience, along with the usual matrix multiplication, we use here the non-standard Hadamard product \cite{Styan:73,Reams:99}, which is the matrix of elementwise products $(A\circ B)_{ij}=A_{ij} B_{ij}$. The Hadamard product is commutative unlike regular matrix multiplication, but the distributive and associative properties are retained. It has previously been proven to be a useful tool when non-scalar mass matrices and non trivial flavor symmetry contractions were involved \cite{Osipov:21plb,Osipov:21prd,Osipov:17}. Matrices $g_\sigma$ and $g_V$ are symmetric and real-valued
\begin{equation}
\label{gv}
(g_\sigma)_{ij}=\frac{1}{\sqrt{4 I_2(M_i,M_j)}}, \quad g_V=\sqrt{6} g_\sigma , 
\end{equation}
where
\begin{equation}
I_2(M_i,M_j)=\frac{-i N_c}{(2\pi)^4}\int_\Lambda \frac{d^4p}{(p^2-M_i^2)(p^2-M_j^2)}=\frac{I_1(M_i)-I_1(M_j)}{M_j^2-M_i^2}.
\end{equation}
In the case of equal masses we have
\begin{equation}
I_2(M_i,M_i)=\frac{N_c}{(4\pi)^2}\left[\ln\left(1+\frac{\Lambda^2}{M_i^2}\right)-\frac{\Lambda^2}{\Lambda^2+M_i^2}\right].
\end{equation}
Details of the regularization we use to calculate quark one-loop integrals can be found in \cite{Osipov:21prd}. In the following, we also use the notation $I_2$, understanding it as a matrix consisting of elements $(I_2)_{ij}=I_2(M_i,M_j)$.

\subsection{$PA$-transitions}
Approaches based on chiral Lagrangians containing vector and axial-vector states in addition to pseudoscalar and scalar fields face the need to eliminate off-diagonal scalar-vector and pseudoscalar-axial-vector ($PA$) transitions \cite{Geffen:69}. This is achieved by appropriately redefining the vector and axial-vector fields. Here we obtain a general form of the axial-vector field transformation that eliminates off-diagonal $PA$ transitions. As we will see below, $PA$ diagonalization is one of the reasons for the appearance of the parameters $\delta$ and $\delta'$ in the $\eta^{(\prime)}\to\pi^+\pi^-\gamma$ decay amplitude.  

From the Lagrangian (\ref{L2}) it follows that the result of separating the divergent part in the integral describing the $P\to A_\mu$ transition via the virtual exchange of a quark-antiquark pair is a local vertex
\begin{equation}
\label{PA1}
L_{PA}=-2\langle \left(I_2\circ  \partial_\mu\phi\right)  \{A_\mu, M \}\rangle. 
\end{equation}
Eq.\,(\ref{PA1}) can be recast into a somewhat more explicit form if we make use of the simple relation $\{M, A_\mu\} = \Sigma_M \circ A_\mu$ which is fulfilled for the diagonal matrix $M$ and where $\Sigma_M$ is a mass-dependent matrix with elements $(\Sigma_M)_{ij}=M_i+M_j$. Considering that the matrices $I_2$ and $\Sigma_M$ are symmetric, and the product of the matrices is under the trace sign, we obtain
\begin{eqnarray}
\label{PA2}
L_{PA}&=&-2\langle \left(I_2\circ  \partial_\mu\phi\right)  \left(\Sigma_M \circ A_\mu\right)\rangle=-2\langle \partial_\mu\phi \left(I_2\circ \Sigma_M \circ A_\mu \right)  \rangle \nonumber \\
&=&-\langle \left(g_\sigma \circ \partial_\mu\phi'\right) \left(I_2\circ \Sigma_M \circ g_V\circ A_\mu' \right)  \rangle =
-\langle  \partial_\mu\phi' \left(I_2\circ \Sigma_M \circ g_\sigma \circ g_V\circ A_\mu' \right)  \rangle \nonumber \\
&=&-\sqrt{\frac{3}{8}}\langle \partial_\mu \phi' \left(\Sigma_M\circ A_\mu'\right) \rangle  =-\sqrt{\frac{3}{8}}\langle \partial_\mu \phi' \{A_\mu',M \}\rangle .
\end{eqnarray}
We have examined in such detail the transition from the original "bare" fields to physical ones in order to illustrate the technique of working with the Hadamard product, which is rarely encountered in physics applications, but is a useful tool for solving the problem studied here.

As is known \cite{Osipov:17}, to eliminate $PA$ mixing (\ref{PA2}) it is necessary to redefine the axial vector field $A_\mu'$, namely
\begin{equation}
\label{RedA}
A_\mu'=A_\mu''+k\circ \Sigma_M\circ D_\mu\phi',
\end{equation}
where $k$ is the matrix that ultimately solves the problem of eliminating $PA$ transitions, and $D_\mu\phi'=\partial_\mu \phi'-ie\mathcal{A_\mu}[Q,\phi']$. The naive ($A_\mu'=A_\mu''+k\circ \Sigma_M\circ \partial_\mu\phi'$) elimination of PA transitions in the $\gamma\to\pi\pi\pi$ amplitude leads to a well-known contradiction with the low-energy result of Wess and Zumino, obtained on the basis of Ward's anomalous chiral identities \cite{Wakamatsu:89}. The naive procedure also violates the $U(1)_{em}$ gauge invariance of the $a_1/f_1\to\gamma\pi\pi$ amplitudes \cite{Osipov:18a,Osipov:18b}. Both shortcomings can be avoided by using a $U(1)_{em}$ gauge-covariant form (\ref{RedA}) \cite{Osipov:20}. In this case, additional triangle quark diagrams arise (see Fig.\,\ref{f1}) at one of the vertices of which a pion-photon pair is created. This, on the one hand, leads to a deviation from the picture of complete VMD hypothesis of photon couplings (the importance of this step was stressed in \cite{Kugo:85}), and on the other -- to the appearance of surface terms \cite{Jackiw:00}. These steps avoid discrepancies with the chiral anomalous Ward's identities in a manner similar to that described long ago by Jackiw \cite{Jackiw:72}. We will discuss these issues in more detail in the Sect.\,\ref{st}.

To find $k$, we should consider the mass part of the free Lagrangian of the axial vector field $A_\mu$, which, according to (\ref{L2}), can be expressed as
\begin{equation}
L_A=\frac{1}{4G_V}\langle A_\mu A^\mu\rangle +\langle \{M, I_2\circ A_\mu\}\{M, A^\mu\}\rangle 
=\frac{1}{4}\langle A_\mu' \left( M_A^2 \circ A^{\mu\prime} \right) \rangle,
\end{equation}
where the mass matrix $M^2_A$ is
\begin{equation}
\label{M2A}
M^2_A=\frac{3}{2G_V} g_\sigma^{\circ 2} +\frac{3}{2}\Sigma_M^{\circ 2}.
\end{equation}
The standard definition of the Hadamard power is used here, i.e. $g_\sigma^{\circ 2} =g_\sigma \circ g_\sigma$.  

Requiring complete compensation of the non-diagonal transitions $PA$ in the sum of Lagrangians $L_{PA}+L_A$, we find 
\begin{equation}
k=\sqrt{\frac{3}{2}}\left(M_A^2\right)^{\circ -1},
\end{equation}
where the Hadamard inverse of matrix $B$ (with $B_{ij} > 0$) is defined by $(B^{\circ -1})_{ij}= 1/B_{ij}$. Thus, the transformation (\ref{RedA}) takes the following form
\begin{equation}
\label{RedA2}
A_\mu'=A_\mu''+ \sqrt{\frac{3}{2}}\left(M_A^2\right)^{\circ -1} \circ \Sigma_M\circ D_\mu\phi'.
\end{equation}

Moreover, the kinetic term of the pseudoscalar fields receives an additional contribution:
\begin{equation}
L_{\phi'}^{kin}=\frac{1}{4}\langle \partial_\mu \phi' \left[ \left({\bf I}_H-\frac{3}{2}\Sigma_M^{\circ 2} \circ (M_A^2)^{\circ -1}\right)\circ
\partial^\mu\phi' \right]\rangle ,
\end{equation}
where ${\bf I}_H$ is $3\times 3$ matrix consisting of ones. This requires additional redefinition of pseudoscalar fields
\begin{equation}
\phi'=Z^{\circ 1/2}\circ\phi'',
\end{equation}
where the Hadamard square root  $(Z^{\circ 1/2})_{ij}= Z_{ij}^{1/2}$, 
\begin{equation}
\label{Z}
Z=\left({\bf I}_H-\frac{3}{2}\Sigma_M^{\circ 2} \circ (M_A^2)^{\circ -1}\right)^{\circ -1} ,
\end{equation}
and correspondingly 
\begin{equation}
\phi=g_\sigma\circ\phi'=g_\sigma\circ Z^{\circ 1/2} \circ \phi''\equiv g_\phi\circ\phi''.
\end{equation}

Thus, in terms of physical fields, the transformation excluding PA mixing has the form
\begin{equation}
\label{RedA3}
A_\mu'=A_\mu''+ \sqrt{\frac{3}{2}}\left(M_A^2\right)^{\circ -1} \circ \Sigma_M\circ Z^{\circ 1/2}\circ D_\mu\phi''.
\end{equation}

Next we apply the Goldberger-Treimann relation at the quark level
\begin{equation}
\label{GTr}
g_\phi=\frac{1}{2}\Sigma_M\circ F_\phi^{\circ -1},
\end{equation}
where $F_\phi$ is the matrix containing a set of physical pseudoscalar meson decay constants (e.g. $(F_\phi)_{12}= (F_\phi)_{ud}=f_\pi=92.2\,\mbox{MeV}$) which enters the PCAC type coupling of the weak meson current to weak gauge bosons. Then, using the Goldberger-Treiman relation (\ref{GTr}) and formulas (\ref{gv}), (\ref{M2A}) and (\ref{Z}) we find
\begin{equation}
\sqrt{\frac{3}{2}}\left(M_A^2\right)^{\circ -1} \circ \Sigma_M\circ Z^{\circ 1/2}=\sqrt 6 \left(M_A^2\right)^{\circ -1} \circ Z \circ g_\sigma\circ F_\phi  =4 G_V g_V^{\circ -1} \circ F_\phi .
\end{equation}
This gives
\begin{equation}
\label{RedA4}
A_\mu'=A_\mu''+ 4 G_V g_V^{\circ -1} \circ F_\phi\circ D_\mu\phi''=A_\mu''+ \left(a_\phi\circ g_V\circ F_\phi \right)^{\circ -1}\circ D_\mu\phi''.
\end{equation}

In the final step, we converted the expression to the form used in \cite{Osipov:2020vad}. This allows us to obtain an explicit expression for the constants $1/a$ and $1/a_\eta$ used there, but now in the form of a symmetric $3\times 3$ matrix, where they represent only two of the six independent elements
\begin{equation}
\label{aphi}
a_\phi^{\circ -1} = 4G_V F_\phi^{\circ 2}=4G_V \left( 
\begin{array}{ccc}
\! f_u^2   & f_{\pi^+}^2 & f_{K^+}^2\! \\
\! f_{\pi^-}^2 & f_d^2 & f_{K^0}^2\!   \\
\! f_{K^-}^2& f_{\bar K^0}^2 & f_s^2 \! \\
\end{array}
\right),
\end{equation}
where the value of the constant $G_V$ can be obtained using the model expression for the $\rho$-meson mass $m^2_\rho=g_\rho^2/(4G_V)$. The decay constant $g_\rho\simeq \sqrt{12\pi}$ is known from the decay width $\rho\to\pi\pi$. From this, in particular, it follows that 
\begin{equation}
\label{a}
a_\pi\equiv a=m^2_\rho/(g_\rho f_\pi)^2.
\end{equation}

If we take into account that the matrices of the pseudoscalar field $\phi''$ and the axial-vector field $A_\mu''$ have the form
\begin{equation}
\phi'' = 
\left( 
\begin{array}{ccc}
\phi_u'' &\sqrt 2 \pi^+& \sqrt 2 K^+ \\
\sqrt 2 \pi^-& \phi_d'' & \sqrt 2 K^0 \\
\sqrt 2 K^- & \sqrt 2 {\bar K}^0 & \phi_s''  
\end{array}\right), \quad
\label{AVf}
A_\mu'' = 
\left( 
\begin{array}{ccc}
A_{u\mu}'' &\sqrt 2 a_{1\mu}^+& \sqrt 2 K_{1\mu}^+ \\
\sqrt 2 a_{1\mu}^-& A_{d\mu}'' & \sqrt 2 K_{1\mu}^0 \\
\sqrt 2 K_{1\mu}^- & \sqrt 2 \bar K_{1\mu}^0 & A_{s\mu}''  
\end{array}\right),
\end{equation}
then it is easy to write out a transformation that eliminates the $\pi a_\mu$ mixing  
\begin{equation}
\label{RedA5}
A_\mu^{\pm \prime}=a_\mu^\pm +\frac{1}{a g_\rho f_\pi} D_\mu \pi^\pm. 
\end{equation}

To obtain a transformation corresponding to a certain axial vector field from the matrix (\ref{AVf}), it is necessary to select an element corresponding to this field in each of the matrices included in the Hadamard product (\ref{RedA4}). In addition to (\ref{RedA5}), we will further need a shift of the axial vector field $f_{1\mu}^{ns}$, leading to vertices with $\eta$ and $\eta'$ mesons. This field is contained in two diagonal elements of the matrix (\ref{AVf})
\begin{eqnarray}
A_{u\mu}'' &=&A_{3\mu}'' +\frac{\sqrt 2 A_{0\mu}''+A_{8\mu}''}{\sqrt 3}=a_{1\mu}^0 + f_{1\mu}^{ns}, \nonumber \\
A_{d\mu}'' &=&-A_{3\mu}'' +\frac{\sqrt 2 A_{0\mu}''+A_{8\mu}''}{\sqrt 3}=-a_{1\mu}^0 + f_{1\mu}^{ns}, \nonumber \\
A_{s\mu}'' &=& \frac{\sqrt 2 A_{0\mu}''-2 A_{8\mu}''}{\sqrt 3}= \sqrt 2 f_{1\mu}^{s}, 
\end{eqnarray}
and is projected from the general formula (\ref{RedA4}) by the operator $P_{f_1^{ns}}=(\sqrt 2 \lambda_0+\lambda_8)/\sqrt 3$. That gives  
\begin{eqnarray}
&&\frac{1}{2}\langle A_{\mu}''P_{f_1^{ns}}\rangle = f_{1\mu}^{ns}, \quad  \frac{1}{2}\langle a_{\phi}^{\circ -1}P_{f_1^{ns}}\rangle = \frac{g_\rho^2}{2 m_\rho^2}\left(f_u^2+f_d^2\right), \nonumber \\
&&\frac{1}{2}\langle  g_V^{\circ -1}P_{f_1^{ns}}\rangle=\frac{1}{2}\left(\frac{1}{g_V(M_u)} +\frac{1}{g_V(M_d)}\right)\simeq \frac{1}{g_\rho}, \nonumber \\
&& \frac{1}{2}\langle D_\mu \tilde\phi P_{f_1^{ns}}\rangle =\partial_\mu \frac{\sqrt 2 \tilde\phi_0 +\tilde \phi_8}{\sqrt 3}.
\end{eqnarray}

Finally, we find a transformation that eliminates the $f_{1\mu}^{ns}$-$\eta^{ns}$ mixing 
\begin{equation}
\label{ns}
f_{1 \mu}^{ns \prime} = f_{1 \mu}^{ns} +\frac{1}{a_{\eta}g_\rho f_\pi} 
\partial_\mu \frac{\sqrt 2 \tilde\phi_0 +\tilde \phi_8}{\sqrt 3}f_\pi 
\end{equation}
where 
\begin{equation}
\label{aeta}
\frac{1}{a_\eta} = \frac{g_\rho^2}{2m_\rho^2}\left(f_u^2+f_d^2 \right)=\frac{1}{a}\left(\frac{f_u^2+f_d^2}{2f_\pi^2}\right).
\end{equation}

It should be noted that the constant $f_\pi\equiv f_{\pi^+}$ is experimentally well determined, which cannot be said about the constant $f_{\pi^0}$. Theoretically, there are two contributions to the difference $f_{\pi^+}-f_{\pi^0}\equiv \Delta f_\pi$. In pure QCD $\Delta f_\pi$ is quadratic in the quark mass difference $(m_d - m_u)$ and is estimated in ChPT to be of order $\Delta f_\pi \simeq 0.7\times 10^{-4} f_{\pi^0}$ \cite{Moussallam:09}. This value is negligibly small. The other contribution is purely electromagnetic and, being of order $O (\alpha_{QED})\simeq 10^{-2}$. Thus, when calculating the difference $(1/a_\eta -1/a)$ in the tree approximation of the NJL model, the electromagnetic contribution will dominate
\begin{equation}
\label{diffa}
\frac{1}{a}-\frac{1}{a_\eta}=\frac{2}{a f_\pi}(\Delta f_\pi)_{QED} + O( (m_d-m_u)^2).
\end{equation}

\subsection{$\eta$-$\eta'$ mixing}
To move to physical fields in formula (\ref{ns}), it is necessary to establish a connection between dimensionless variables $\tilde\phi_i$ and dimensional ones $\phi'$ and then take into account the mixing of $\eta$-$\eta'$ fields. Recall that for the diagonal elements we have 
\begin{equation}
\label{rescaling}
      f_i \tilde\phi_i =\phi''_i, \quad (i=u,d,s).
\end{equation}
To move from the flavor basis $\tilde\phi_i$ to the singlet-octet $\phi_a$ $(a=0,3,8)$, we use the result of  \cite{Osipov:23b}, where it is shown that  
\begin{eqnarray}
\label{mixphi}
\tilde\phi_0 &=& \frac{\phi_0''}{f_0}\!+\! \left(\frac{1}{f_u}\!-\!\frac{1}{f_d} \right) \frac{\phi_3''}{\sqrt 6}  
\!+\! \left(\frac{1}{f_u}\!+\!\frac{1}{f_d}\!-\!\frac{2}{f_s}\right) \frac{\phi_8''}{3\sqrt 2},  \nonumber \\            
\tilde\phi_8 &=& \frac{\phi_8''}{f_8} \!+\! \left(\frac{1}{f_u}\!-\!\frac{1}{f_d} \right) \frac{\phi_3''}{2\sqrt 3} 
\!+\!\left(\frac{1}{f_u}\!+\!\frac{1}{f_d}\!-\!\frac{2}{f_s}\right) \frac{\phi_0''}{3\sqrt 2}, \nonumber\\
\tilde\phi_3 &=& \frac{\phi_3''}{f_3} \!+\! \left(\frac{1}{f_u}\!-\!\frac{1}{f_d}\right)
\frac{\phi_8'' \!+\!\sqrt 2 \phi_0''}{2\sqrt 3},               
\end{eqnarray}
where 
\begin{equation}
\label{f083}
\frac{1}{f_0}=\frac{1}{3} \left(\frac{1}{f_u} \!+\!\frac{1}{f_d}\! +\!\frac{1}{f_s}\right), \quad
\frac{1}{f_8}=\frac{1}{6} \left( \frac{1}{f_u} \!+\! \frac{1}{f_d}\! +\!\frac{4}{f_s} \right),  \quad
\frac{1}{f_3}=\frac{1}{2} \left( \frac{1}{f_u}\!+\!\frac{1}{f_d} \right). 
\end{equation}

The physical fields are the result of an orthogonal transformation that diagonalizes the mass matrix of $\phi_0''$, $\phi_8''$ and $\phi_3''$ states. To first order in isospin breaking, one can use the transformation \cite{Leutwyler:96}
\begin{equation}
\label{RtoPh}
\left( 
\begin{array}{c}
\phi_0'' \\ \phi_8''\\ \phi_3'' \\
\end{array}
\right)
=U(\theta, \epsilon, \epsilon' )
\left( 
\begin{array}{c}
\! \eta' \! \\ \! \eta \! \\ \! \pi^0 \!  \\
\end{array}
\right).
\end{equation}
where $U$ is defined by
\begin{equation}
\label{U}
U(\theta, \epsilon, \epsilon' )=\left( 
\begin{array}{ccc}
\! \cos\theta & -\sin\theta & \epsilon'\cos\theta\!-\!\epsilon \sin\theta\!    \\
\! \sin\theta  & \cos\theta & \epsilon' \sin\theta\!+\!\epsilon \cos\theta\! \\
\! -\epsilon' & -\epsilon & 1 \!\\
\end{array}
\right), \quad UU^{-1}={\bf 1}+ O ((m_d-m_u)^2).
\end{equation}
The matrix is parametrized by three angles $\theta$, $\epsilon$, $\epsilon'$. The first arises from the mass difference of the strange and nonstrange quarks and breaks $SU(3)$, i.e., in the limit of exact $SU(3)$ symmetry $\theta\to 0$. The other two angles describe the isospin breaking effects. They are proportional to the difference $m_d-m_u$. Since we consider here only the decays of $\eta$ and $\eta'$ mesons, the part related to the neutral pion field is discarded in what follows.

There are several points worth noting here. 

(a) Using formulas (\ref{mixphi}), we find that the NJL model leads to a picture with one mixing angle:
\begin{equation}
\label{cetaNJL}
 \frac{\sqrt 2 \tilde\phi_0 +\tilde \phi_8}{\sqrt 3} f_\pi =\frac{f_\pi}{f_3\sqrt 3}\left[\eta \left(\cos\theta-\sqrt 2\sin\theta \right) + \eta' \left(\sqrt 2 \cos\theta+\sin\theta \right)\right]+\ldots ,
\end{equation}
where the ellipsis denotes the contribution of the $\pi^0$ meson, which is not significant for our further consideration. This is an oversimplified picture which, as is well known, is not able to describe, for example, all three two-photon decays $\pi^0, \eta, \eta'\to \gamma\gamma$.

(b) A more realistic picture emerges if the terms in formulas (\ref{mixphi}) are classified in accord with the $1/N_c$ expansion. In this case, the first term is the leading one in all three expressions. The contribution of the remaining terms is suppressed as $1/N_c$  \cite{Osipov:23b,Osipov:23a}. Neglecting these contributions, we obtain
\begin{equation}
 \frac{\sqrt 2 \tilde\phi_0 +\tilde \phi_8}{\sqrt 3} f_\pi =  c_\eta \eta + c_{\eta'}\eta' +\ldots ,
\end{equation}
where the coefficients $c_{\eta^{(\prime )}}$ are  
\begin{eqnarray}
\label{cetacoeff}
    c_\eta & = & \frac{1}{\sqrt 3}\left(\frac{f_\pi}{f_8}  \cos{\theta}  - \sqrt 2 \frac{f_\pi}{f_0} \sin{\theta}\right), \nonumber\\
    c_{\eta'} & = &\frac{1}{\sqrt 3}\left( \frac{f_\pi}{f_8}  \sin{\theta}  + \sqrt 2 \frac{f_\pi}{f_0} \cos {\theta}\right).
\end{eqnarray}
Using the experimental data on the two-photon decay widths, and the ratio $f_8/f_\pi \simeq 1.30$ from the $SU(3)$ ChPT one can obtain that $\theta \simeq -20^\circ$, and $f_0/f_\pi \simeq 1.04$ \cite{Gasser:84,Gasser:85}. This parameterization is shown to be consistent with anomalous $\eta/\eta' \to\pi^+\pi^-\gamma$  decays \cite{Holstein:98}. A careful study of final state interactions and unitarity constraints in order to realistically extrapolate to zero four-momenta, as required by the anomaly, allowed Venugopal and Holstein to establish the phenomenological values of these parameters and show that they are consistent with those obtained from ChPT. This scheme is known as the  parameterization in terms of two (octet $f_8$ and singlet $f_0$) decay constants and one $\eta$-$\eta'$ mixing angle $\theta$.

\subsection{Surface terms in the $\eta^{(\prime)}\to\pi^+\pi^-\gamma$ amplitude}
\label{st}

A peculiarity of the $\eta^{(\prime)}\to\pi^+\pi^-\gamma$ amplitude is that there are two types of $PA$ transitions: $\pi$-$ a_1$ (\ref{RedA5}) and $\eta^{(\prime)}$-$f_1^{ns}$ (\ref{ns}). This results in a nonzero correction to the contact VAAA anomaly, which vanishes in the chiral limit and therefore does not conflict with the chiral Ward identities. It is this correction that leads to the appearance of the parameter $\delta^{(\prime )}$ in the amplitude. For the sake of completeness, we will present the details of this mechanism here. For this purpose, let us write out the $\eta^{(\prime)} \to\pi^+\pi^-\gamma$ decay amplitude,
\begin{eqnarray}
\label{ampl}
A_{\eta^{(\prime)}\to\pi\pi\gamma}&=&A_{\mbox{\scriptsize box}}+A_{\rho} + A_{\mbox{\scriptsize st}}
=\frac{e N_c c_{\eta^{(\prime)}}}{12 \pi^{2} f_{\pi}^{3}}  e_{\mu\nu\alpha\beta}\, \epsilon^{*\mu}(p_{\gamma})\, p_{\gamma}^\nu p_{+}^\alpha p_{-}^\beta \nonumber \\
&\times& \left[1 + \frac{1}{a} - \frac{1}{a_\eta}+\frac{1}{a}\left(\frac{2}{a_\eta}-\frac{1}{2a}\right) +\frac{1-12 b}{8a^2a_\eta} +\left(\frac{3}{a}\right)\frac{q^2}{m_\rho^2-q^2}\right].
\end{eqnarray}
It includes three types of contributions: the box diagram contribution $A_{\mbox{\scriptsize box}}$, the resonance contribution $A_{\rho}$ describing the production of a $\pi^+\pi^-$ pair due to the $\rho$-meson decay, and the contribution of the surface term $A_{\mbox{\scriptsize st}}$. The following notations are used: $p_{\gamma}$  and $\epsilon^{\mu}(p_{\gamma})$ are the four-momentum and polarization of the photon, $p_+, p_-$ are the momenta of the pions, and $q=(p_+ +p_-)$. This expression is obtained on the basis of the Lagrangian (\ref{L2}), where the imaginary (anomalous) part of the fermion determinant is obtained by isolating the leading part of the quark loops in the expansion in momenta.   

The constant $b$ in the square brackets of (\ref{ampl}) appears as a result of taking into account the surface term, which is associated with the second diagram shown in Fig.\,\ref{f1}. Notice that the first diagram corresponds to the case of pseudoscalar interaction of the pion with quarks $\bar q\gamma_5\pi q$, and the second to pseudovector interaction $\bar q \gamma_\mu \gamma_5 \partial^\mu \pi q$. It is easy to establish that the contribution of the first diagram is zero. The contribution of the second diagram would also be zero if not for the shift ambiguity associated with the formal linear divergence of the integrals arising here.

\begin{figure*}[t]
 \centering
  \begin{subfigure}{0.5\textwidth}
   \centering
   \begin{tikzpicture}
    \begin{feynman} 
      \vertex (l) {\( \eta^{(\prime)} \)};
      \vertex [dot, right=1.2cm of l] (a) {};
      \vertex [dot, right=1.2cm of a] (b) {};
      \vertex [dot, above right=of b] (c) {};
      \vertex [dot, below right=of b] (d) {};
      \vertex[dot, right=1.0cm of c] (e) {};
      \vertex [right=1.3cm of d] (h) {\( \pi^{\pm} \)};
      \vertex [dot, right=1.0cm of c] (e) {};
      \vertex [above right=1.0cm of e] (f) {\( \gamma \)};
      \vertex [below right=1.0cm of e] (g) {\(  \pi^{\mp} \)};
      \diagram* {
        (l) -- [scalar] (a),
        (a) -- [double, edge label'=\( f_1^{ns} \)] (b),
        (b) -- [anti fermion] (c),
        (c) -- [anti fermion] (d),
        (d) -- [anti fermion] (b),  
        (d) -- [scalar] (h),
        (c) -- [double, edge label'=\( a_1^\mp \)] (e),
        (e) -- [photon] (f), 
        (e) -- [scalar] (g), 
      };
     \end{feynman}
    \end{tikzpicture}
  \end{subfigure}%
 \centering
  \begin{subfigure}{0.5\textwidth}
   \centering
   \begin{tikzpicture}
    \begin{feynman} 
      \vertex (l) {\( \eta^{(\prime)} \)};
      \vertex [dot, right=1.2cm of l] (a) {};
      \vertex [dot, right=1.2cm of a] (b) {};
      \vertex [dot, above right=of b] (c) {};
      \vertex [dot, below right=of b] (d) {};
      \vertex[dot, right=1.0cm of d] (e) {};
      \vertex [right=1.0cm of e] (h) {\( \pi^{\pm} \)};
      \vertex [dot, right=1.0cm of c] (k) {};
      \vertex [above right=1.0cm of k] (f) {\( \gamma \)};
       \vertex [below right=1.0cm of k] (g) {\( \pi^{\mp} \)};
      \diagram* {
         (l) -- [scalar] (a),
        (a) -- [double, edge label'=\( f_1^{ns} \)] (b),
        (b) -- [anti fermion] (c),
        (c) -- [anti fermion] (d),
        (d) -- [anti fermion] (b),  
        (d) -- [double, edge label'=\( a_1^\pm \)] (e),
        (e) -- [scalar] (h),
        (c) -- [double, edge label'=\( a_1^\mp \)] (k),
        (k) -- [scalar] (g),
        (k) -- [photon] (f), 
      };
     \end{feynman}
    \end{tikzpicture}
  \end{subfigure}%
 \caption{Additional quark triangle diagrams arising in the $\eta^{(\prime)}\to \pi^+\pi^-\gamma$ decay amplitude from the substitution (\ref{RedA4}) with two (left panel) and three (right panel) $PA$ transitions. Both diagrams do not describe the propagation of the physical $a_1^\pm$ and $f_1^{ns}$ mesons, but should be understood as a graphical representation of the contribution from the $\pi$-$a_1$ and $\eta^{(\prime)}$-$f_1^{ns}$ mixing.}
 \label{f1}
\end{figure*}%

Let us examine this subtle point in more detail. To do this, we will write out the loop-integrals corresponding to the second diagram.
\begin{equation}
\label{pov}
T^\beta (l,p_+,p_-)= \left[J_1^{\beta\alpha}(l,p_-) 
-  J_2^{\beta\alpha}(l,p_-)+ J_1^{\beta\alpha}(l,p_+) -J_2^{\beta\alpha}(l,p_+) \right] \epsilon^*_\alpha (p_\gamma) ,
\end{equation}
where the index $\beta$ is further summed with the $\eta^{(\prime)}$ meson momentum $l_\beta$. The integrals are 
\begin{eqnarray}
\label{J1}
J_1^{\beta\alpha}(l,p_-)&=&\!\!\int\!\! \frac{d^4k}{(2\pi
                          )^4}\mbox{tr}\left[ S(k,0)
                          \gamma^\beta\gamma_5  S(k,l)  \hat p_-
                          \gamma_5 S(k,l-p_- )\gamma^\alpha\gamma_5 \right],      \\
 \label{J2} 
J_2^{\beta\alpha}(l,p_-)&=&\!\!\int\!\! \frac{d^4k}{(2\pi
                          )^4}\mbox{tr}\left[ S(k,0)
                          \gamma^\beta\gamma_5  S(k,l) \gamma^\alpha\gamma_5 
                         S(k,p_- )\hat p_-\gamma_5\right],      \\
S(k,l)&=& \frac{\hat k-\hat l +M}{(k-l)^2-M^2}, 
\end{eqnarray}
and $\hat k=k_\mu\gamma^\mu$. The subtlety here is that if it were possible to shift the integration variable, then the first term would completely cancel the second, and the third term would cancel the fourth term in square brackets (\ref{pov}), and we would obtain $T^\beta = 0$ (This cancellation led to the vanishing of the contribution of the first diagram in Fig.\,\ref{f1}). However, as is well known \cite{Jackiw:00}, due to the formal linear divergence of the integrals, which is present
in (\ref {J1}) and (\ref{J2}) even after calculating the traces, surface terms arise, making the result nonzero.
\begin{equation}
 J_1^{\beta\alpha}(l,p_-)-J_2^{\beta\alpha}(l,p_-)=\frac{1}{8\pi^2}
e^{\mu\nu\alpha\beta} c_\mu  (p_-)_\nu ,
\end{equation} 
where $c_\mu$ is an arbitrary four-vector. Accordingly, for the whole sum (\ref{pov}) we find
\begin{equation}
\label{povterm}
  T^\beta (l,p_+,p_-)=\frac{1}{8\pi^2}  \epsilon^*_\alpha (p_\gamma) e^{\mu\nu\alpha\beta} c_\mu (p_+ + p_-)_\nu .
\end{equation}

The vector $c^\mu$ can be represented as a linear combination of three independent vectors that directly related to the process under consideration $c^\mu = \tilde a p_\gamma^\mu + b(p_+ - p_-)^\mu + \tilde c (p_+ + p_-)^\mu $. In this case, only the first two survive after substituting this expression into (\ref{povterm}). As a result, we obtain for (\ref{povterm})  
\begin{equation}
\label{pt}
T^\beta (l,p_+,p_-)=\frac{1}{8\pi^2}  \epsilon^*_\alpha (p_\gamma) e^{\alpha\beta}_{\ \ \ \, \mu\nu} 
\left[\tilde a p_\gamma^\mu (p_+ + p_-)^\nu +2b p_+^\mu p_-^\nu \right].
\end{equation}
Considering that after multiplying by $p_{\eta^{(\prime)}}^\beta =(p_++p_-+p_\gamma)^\beta $ the first term in (\ref{pt}) becomes zero, we conclude that only the constant $b$ finally remains in the expression for the amplitude (\ref{ampl}). This coupling can be uniquely fixed in accord with the low-energy theorem, namely $b=a+1/12$. Thus, we finally obtain 
\begin{equation}
\label{tot1}
A_{\eta^{(\prime)}\to\pi\pi\gamma}=\frac{e N_c c_{\eta^{(\prime)} }}{12 \pi^{2} f_{\pi}^{3}}  \left[1 +\delta^{(\prime)} +\left(\frac{3}{a}\right)\frac{q^2}{m_\rho^2-q^2}\right] e_{\mu\nu\alpha\beta}\, \epsilon^{*\mu}(p_{\gamma})\, p_{\gamma}^\nu p_{+}^\alpha p_{-}^\beta ,
\end{equation}
where
\begin{equation}
\label{delta}
\delta^{(\prime)}=\left(\frac{1}{a}-\frac{1}{a_\eta}\right)\left(1-\frac{1}{2a}\right).
\end{equation}

Thus, we can conclude that the parameter $\delta^{(\prime)}$ is a residual effect when eliminating $PA$-transitions that violate the anomalous Ward identities.

\section{General features of the approach used}

Let us discuss the main assumptions behind our further calculations.

{\it 1. Amplitude.} The analysis will use the amplitude (\ref{tot1}), in which we set
\begin{equation}
\label{amp1}
A(\eta^{(\prime)}\to\pi^+\pi^-\gamma)=A_{\eta^{(\prime)}}\left(1+\delta^{(\prime)}+\frac{3}{a}\ \frac{s_{\pi\pi}}{m^2_\rho-s_{\pi\pi}}\right)
\end{equation}
with $A_{\eta^{(\prime )}}$ defined in (\ref{A}). When $\delta^{(\prime)}=0$ and $a=2$ it coincides with the result of the hidden local symmetry approach \cite{Picciotto:92}. 

Although we used the NJL model in the previous section to obtain this amplitude, it is worth noting several significant points in which our analysis deviates from the standard NJL consideration.

(a) The NJL model describes perfectly the $\pi^0\to\gamma\gamma$ decay. However, to take into account interactions involving $\eta$ or $\eta'$ mesons (for instance, $\eta^{(\prime)}\to\gamma\gamma$ decays), the effects of $SU(3)$ and $U(3)$ symmetry breaking must be taken into account. As we showed above, breaking these symmetries at the level of quark loops leads to a single-angle mixing picture, which, however, does not include the important singlet, $f_0$, and octet, $f_8$, constants. These effects are absent in formula (\ref{cetaNJL}), since they are associated with the inclusion of meson loops, which are beyond the accuracy of the NJL model. As we have already noted, the problem can be solved within the framework of the $1/N_c$ expansion. Steps in this direction are already being taken \cite{Osipov:23a,Osipov:23b}, but the problem still remains open. Therefore, in our estimates of the constants $c_{\eta^{(\prime )}}$, we turn to the well-known result of ChPT (\ref{cetacoeff}).

(b) The NJL model, although allowing us to conclude that the amplitude (\ref{tot1}) contains the parameter $\delta^{(\prime)}$, underestimates its magnitude (from Eqs.\,(\ref{diffa}) and (\ref{delta}) it follows that $\delta^{(\prime)}\sim 10^{-2}$). The reason is also the lack of contributions from meson one-loop diagrams. This is why in our subsequent calculations we leave this parameter free and determine it based on experimental data. There are two independent ways to find the value of $\delta^{(\prime)}$. The first is to determine $\delta^{(\prime)}$ based on empirical partial width data 
\begin{eqnarray}
\label{expeta}
\Gamma^{exp}_{\eta\to\pi^+\pi^-\gamma}&=&56.07\pm 3.03\,\mbox{eV}\ \, \quad \mbox{\cite{PDG:24}}, \\
\label{expetap}
\Gamma^{exp}_{\eta^\prime\to\pi^+\pi^-\gamma}&=&56.21\pm 2.87\,\mbox{keV} \quad \mbox{\cite{BESIII:19}} .
\end{eqnarray}
 In this case, details of the $\eta$-$\eta^\prime$ mixing mechanism are also important. This is what was done in \cite{Osipov:2020vad}, where we used the parameterization in terms of two (octet and singlet) decay constants and one $\eta$-$\eta'$ mixing angle. It gives that $\delta=-0.1$, and $\delta^\prime=-0.3$. The second way is to measure the dipion invariant mass distribution and perform a fit using a model-independent approach. It does not require information about the $\eta$-$\eta^\prime$ mixing parameters and allows one to extract the value of another low-energy parameter -- the slope $\alpha^{(\prime )}$. This is what was done, for instance, by WASA-at-COSY \cite{WASA:2012} and KLOE \cite{KLOE:2013} Collaborations to get (\ref{WASA}) and (\ref{KLOE}). As we will show below (see Fig.\,\ref{f4}), differential distributions do not allow us to extract information about the values of $\delta^{(\prime)}$, since they are practically insensitive to it.
 
{\it 2. Slope parameter.} An important consequence of formula (\ref{tot1}) is that the use of the vector dominance hypothesis (a direct consequence of the inclusion of vector mesons in the NJL model) allows us to express the slope parameter $\alpha^{(\prime )}$ through the remaining phenomenological parameters. Indeed, with the help of simple algebra
\begin{equation}
1+\delta^{(\prime)}+\frac{3}{a}\ \frac{s_{\pi\pi}}{m^2_\rho-s_{\pi\pi}}=\left[\left(1-\frac{3}{a} +\delta^{(\prime)}\right) \left( 1-\frac{s_{\pi\pi}}{m^2_\rho}  \right)+\frac{3}{a}\right]\frac{m^2_\rho}{m^2_\rho-s_{\pi\pi}},
\end{equation}
the expression (\ref{amp1}) can be easily reduced to the form (\ref{Gampl}) 
\begin{equation}
\label{amp2}
A(\eta^{(\prime)}\to\pi^+\pi^-\gamma)=A_{\eta^{(\prime )}} \left(1+\delta^{(\prime )}\right) \left(1+\alpha^{(\prime )} s_{\pi\pi} \right) 
\frac{m_\rho^2}{m^2_\rho-s_{\pi\pi}},
\end{equation}
where, however, the parameters $\delta^{(\prime )}$ and $\alpha^{(\prime )}$ are not independent, but are related by the formula 
\begin{equation}
\label{alpha2}
\alpha^{(\prime)}=\frac{1}{m^2_\rho}\left[\frac{3}{a(1+\delta^{(\prime)})}-1\right]=\frac{1}{2m^2_\rho}\left[1-\frac{6\delta^{(\prime)}}{a(1+\delta^{(\prime)})}+\frac{3}{a}(2-a)\right]=\frac{1}{12}\langle r^2\rangle +\tilde\alpha^{(\prime)}.
\end{equation} 

Let us make two observations. 

(a) The first concerns the physical content of the formula (\ref{alpha2}). In this expression, the first term is the mean square of the pion charge radius $\langle r^2\rangle_{\pi^\pm}=6/m^2_\rho\simeq 0.39\,\mbox{fm}^2$. This VMD estimate is somewhat lower ($\sim 10\%$) than the known results of phenomenological analysis (see, for example, $\langle r^2\rangle_{\pi^\pm}=0.437(3)\,\mbox{fm}^2$ \cite{Meissner:12}), but corresponds to the standard $10\%-20\%$ accuracy inherent in the VMD model. The two remaining terms describe the contribution from the explicit $SU(3)$ symmetry breaking by the surface term $\delta^{(\prime)}$ and the contribution violating the KSFR (Kawarabayashi-Suzuki-Fayyazuddin-Riazuddin) relation \cite{KS:66,FR:66}. Both of the above effects are combined into a single parameter $\tilde\alpha^{(\prime )}$, which is analogous to the separation of contributions in \cite{Meissner:12}. If we strictly follow the Ward identities and the KSFR relation $a=2$, as assumed in \cite{Picciotto:92}, then it is obvious that $\tilde \alpha^{(\prime )}=0$ and $\alpha^{(\prime)}=0.83\,\mbox{GeV}^{-2}$. This is significantly lower than the experimental values (\ref{WASA}), (\ref{KLOE}) and (\ref{C-Bc}). Considering that the KSFR relation is satisfied quite well: $a=2.0(2)$, we can conclude that the role of the surface term $\delta^{(\prime)}$ in the slope parameter $\alpha^{(\prime)}$ is significant. 

We note that both effects leading to $\tilde\alpha^{(\prime )}\neq 0$ are responsible for the destroying the balance between the VMD triangle anomaly term and the box anomaly (contact term) involving a photon and three pseudoscalars, established by Cohen \cite{Cohen:1989} based on the analysis of anomalous Ward's identities and the KSFR relation. To demonstrate this let us rewrite the amplitude in the form   
\begin{equation}
A(\eta^{(\prime )}\to\pi^+\pi^-\gamma)=A_{\eta^{(\prime )}}\left[c^{(\prime)}_{cont}+c^{(\prime )}_\rho F_V(s_{\pi\pi})\right],   
\end{equation}
where $c^{(\prime )}_{cont} =1+\delta^{(\prime)}-3/a$ and $c^{(\prime)}_\rho =3/a$. It is clear that when $\delta^{(\prime )}=0$ and $a=2$ the weights $c^{(\prime)}_{cont}$, $c^{(\prime )}_\rho$ take the values established by Cohen: $c^{(\prime)}_{cont}=-1/2$, $c^{(\prime )}_\rho =3/2$. On the contrary, the amplitude (\ref{amp2}) yields $c_{cont}=-0.7$, $c_{cont}'=-0.9$ and $c_\rho=c_\rho'=1.6$. In particular, this implies that the surface term $\delta^{(\prime)}$ only affects the weight of the contact term.

\begin{figure}
\includegraphics[height=0.35\textwidth, width=0.6\textwidth]{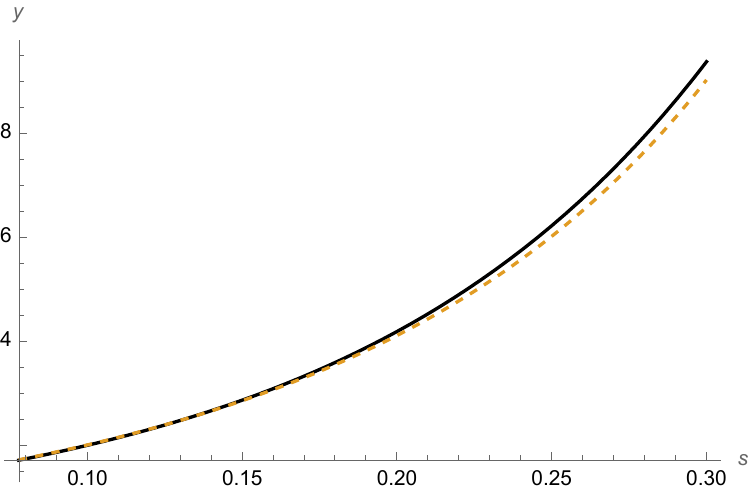}
\caption{The squared modulus $y=|(1+\alpha s_{\pi\pi})F_V(s_{\pi\pi})|^2$ is plotted as a function of $s_{\pi\pi}$ [GeV$^2$]. The solid curve is the result of fitting the slope $\alpha$ to the spectral data of the KLOE Collaboration \cite{KLOE:2013} and corresponds to the central value $\alpha =1.32\,\mbox{GeV}^{-2}$. The pion vector form factor  is well known from direct $e^+ e^-\to\pi^+\pi^-$ measurements and in the region of the $\eta\to\pi^+\pi^-\gamma$ decay
was approximated by a third-order polynomial $|F_V (s_{\pi\pi})|\simeq 1+(2.12\pm 0.01)s_{\pi\pi} +(2.13\pm 0.01)s_{\pi\pi}^2 +(13.80\pm 0.14)s_{\pi\pi}^3 $ \cite{Meissner:12}. The dotted curve corresponds to the value $\alpha= 1.80\,\mbox{GeV}^{-2}$, the VMD form factor (\ref{vmd}) with an imaginary part (\ref{width}).}    
\label{f2}      
\end{figure}

(b) The second remark concerns the nonperturbative part of the amplitude. To take into account unitarity effects in the amplitude (\ref{amp2}) we include the (energy-dependent) width of the $\rho$-meson in the propagator in the standard vector-dominance form
\begin{equation}
\label{vmd}
\frac{m_{\rho}^2}{m^2_\rho -s_{\pi\pi}} \to \frac{m^2_{\rho}}{m_\rho^2-s_{\pi\pi}- i m_\rho \Gamma_{\rho}(s_{\pi\pi})}=F_V(s_{\pi\pi}).  
\end{equation}
 For $s_{\pi\pi}>4m^2_\pi$, the imaginary part of the form factor is nonzero and is characterized by the energy-dependent width of the $\rho\to\pi\pi$ decay
\begin{equation}
\label{width}
\Gamma_{\rho}(s_{\pi\pi})=\frac{g_\rho^2(s_{\pi\pi} -4m_\pi^2)^{3/2}}{48\pi m^2_\rho}\left(\frac{m_\rho}{\sqrt s_{\pi\pi}}\right)^\lambda \theta (s_{\pi\pi}-4m_\pi^2).
\end{equation}
The parameter $\lambda$ is the damping coefficient responsible for the decay of the $\rho$-meson peak. For $\lambda \simeq 2$, the function $|F_V(s_{\pi\pi})|$ reaches its maximum at the point $ s_{\pi\pi}\simeq m_\rho^2$. In the $\eta'\to\pi^+\pi^-\gamma$ decay, the physical region extends over the interval $4m_\pi^2\leq s_{\pi\pi} \leq m_{\eta'}^2$, which almost completely includes the $\rho$ meson peak. In the $\eta\to\pi^+\pi^-\gamma$ decay, the pole singularity lies outside the physical region. The amplitude (\ref{amp2}) provides an oversimplified (VMD) description of the form factor $F_V(s_{\pi\pi})$. On the contrary, the analysis of spectral data requires precise knowledge of its behavior in the entire physical region $4m^2_\pi\leq s_{\pi\pi}\leq m_\eta^2$. In Fig.\,\ref{f2}, we compare the results of the spectral data fitting by the KLOE Collaboration and a similar fitting based on the amplitude (\ref{amp2}). As is clearly seen, the VMD shape of the vector form factor significantly affects the extracted value of $\alpha$. Instead of the value $\alpha = 1.32\,\mbox{GeV}^2$ obtained using the phenomenologically justified vector form factor, the VMD approach leads to a significantly higher value $\alpha = 1.80\,\mbox{GeV}^2$. This indicates the impossibility of simultaneously describing the decay width and the experimentally measured shape of the differential distribution while remaining within the framework of the VMD model, which was emphasized in \cite{Meissner:12}.

{\it 3. $\eta$-$\eta'$ mixing schemes.} In what follows, we will consider mainly the one-angle mixing scheme (\ref{cetacoeff}). This mixing scheme works well within the approach presented here. Adjusting the $\eta/\eta' \to\pi^+\pi^-\gamma$  decay widths to their experimental values (\ref{expeta}) and (\ref{expetap}) enables us to establish that
\begin{eqnarray}
\label{deta}
\delta &=&-0.15(4), \qquad \ \ \alpha=1.46(15)\,\mbox{GeV}^{-2}, \\
\label{detap}
\delta'&=& -0.39^{+0.18}_{-0.20},  \qquad \alpha'=2.71^{-1.02}_{+2.20}\,\mbox{GeV}^{-2}. 
\end{eqnarray} 
Errors in numerical estimates (\ref{deta}) and (\ref{detap}) take into account only error bars with which the decay widths (\ref{expeta}) and (\ref{expetap}) are measured. In the $\eta\to\pi^+\pi^-\gamma$ case, the central value of $\alpha$ is slightly higher than the KLOE Collaboration result (\ref{KLOE}). However, within the margin of error, both results do not contradict each other (the difference is $0.7\sigma$). There is also agreement with the result \cite{Kubis:21}, which states $\alpha=1.52(6)\,\mbox{GeV}^{-2}$. Interestingly, after taking into account the left-hand cut induced by the tensor $a_2(1320)$ state, this result changes only marginally giving $\alpha = 1.42(6)\,\mbox{GeV}^{-2}$ \cite{Kubis:21}, that practically coincides with (\ref{deta}).

It should also be noted that the errors in the slope parameter $\alpha'$ (\ref{detap}) are considerable. Its value changes significantly with a relatively small change in the $\eta'\to\pi^+\pi^-\gamma$ decay width. We see this as an indication that using a first-order polynomial $P_{\eta'}(s_{\pi\pi})$ is not enough and that the subsequent term quadratic in $s_{\pi\pi}$ must be included. High-statistic data also point in favor of such a parameterization \cite{Kubis:17,Ablikim:18}. This seems to indicate the need for more thorough analysis, including the contribution of the $a_2(1320)$ meson. It is known \cite{Kubis:15} that this state significantly distorts the linear behavior of $P_{\eta^{(\prime )}}(s_{\pi\pi})$ beyond $s_{\pi\pi} \leq m_\eta^2$. Recognizing the importance of this step, we plan to address this issue elsewhere.

While the one-angle mixing scheme, although with large error bars, is still suitable for extracting information about the slope $\alpha'$, this cannot be said about the more modern parameterizations in terms of two octet-singlet decay constants and two mixing angles, which follow from the defining matrix elements of the octet and singlet axial-vector current \cite{Leutwyler:98}. Of course, this is not a problem with the parameterization itself. It is primarily related to the structure of the $\eta'\to\pi^+\pi^-\gamma$ amplitude, which, as we have just demonstrated, requires more detailed study in the region $s_{\pi\pi}\geq m^2_\eta$. In the case of $\eta\to\pi^+\pi^-\gamma$ decay, where a linear approximation for the polynomial part of the amplitude is sufficient, no problems arise. 

Table\,\ref{table1} presents the results for the values of the parameters $\delta$ and $\alpha$ that we obtain for various sets of the parameters describing the $\eta$-$\eta'$ system in terms of two angles $(\theta_8, \theta_0)$ and two couplings $(f_8,f_0)$ \cite{Schechter:93,Feldmann:00,Escribano:05,Escribano:16}. This implies that 
\begin{eqnarray}
    c_\eta & = & \frac{1}{\sqrt 3 \cos{(\theta_8-\theta_0)}} \left(\frac{f_\pi}{f_8} 
    \cos{\theta_0}  - \sqrt 2 \frac{f_\pi}{f_0} \sin{\theta_8}\right), \nonumber\\
    c_{\eta'} & = & \frac{1}{\sqrt 3 \cos{(\theta_8-\theta_0)}} \left(\frac{f_\pi}{f_8} 
    \sin{\theta_0}  + \sqrt 2 \frac{f_\pi}{f_0} \cos{\theta_8}\right).
\end{eqnarray}

\begin{table}[tbp!]
\caption{Predictions for the parameters $\delta$ and $\alpha$, obtained in the NJL model by adjusting to the decay width $\Gamma^{exp}_{\eta\to\pi^+\pi^-\gamma}=56.07\pm 3.03\,\mbox{eV}$ \cite{PDG:24} in a scheme with two octet-singlet decay constants and two mixing angles.}
\begin{center}
\small
\begin{tabular}{ccccccc}
\hline
\hline
Ref.
& $f_8/f_\pi$ 
& $f_0/f_\pi$ 
& $\theta_8$
& $\theta_0$
& $\delta$
& $\alpha [\mbox{GeV}^{-2}]$
\\
\hline
\cite{Escribano:16}    
& $1.27(2)$ 
& $1.14(5)$ 
& $-21.2(1.9)^\circ$  
& $-6.9(2.4)^\circ$
& $-0.22\pm 0.04$
& $1.75^{+0.19}_{-0.15}$
\\
\cite{Escribano:05}    
& $1.51(5)$ 
& $1.29(4)$ 
& $-23.8(1.4)^\circ$  
& $-2.4(1.9)^\circ$
& $-0.12\pm 0.04$
& $1.37^{+0.15}_{-0.13}$
\\
\cite{Goity:02}    
& $1.31$ 
& $1.24$ 
& $-20.0^\circ$  
& $-1.5^\circ$
& $-0.16\pm 0.04$
& $1.51^{+0.16}_{-0.14}$
\\
\cite{Roig:25}    
& $1.26(4)$ 
& $1.17(3)$ 
& $-21.2(1.6)^\circ$  
& $-9.2(1.7)^\circ$
& $-0.20\pm 0.04$
& $1.65^{+0.17}_{-0.14}$
\\
\hline
\hline
\end{tabular}
\label{table1}
\end{center}
\end{table}

It will probably be appropriate to make a few comments on these results as well.

Let us start with the general remark regarding the two-angle scheme. In this scheme, the $\eta$-$\eta'$ mixing is reinterpreted in such a way as to be compatible with the large-$N_c$ ChPT at next-to-leading order. This is a more sophisticated approach to the mixing problem, taking into account, in particular, the effects associated with the violation of the OZI rule. Information on the mixing parameters is usually obtained by studying two-photon decays of $\pi^0$, $\eta$ and $\eta'$ mesons, as well as by investigating transition form factor data for single and double photon virtuality. In determining $\delta$, we used the results of such estimates \cite{Escribano:16,Escribano:05,Goity:02,Roig:25}.

Within the specified uncertainties, all sets of parameters, $f_8/f_\pi$, $f_0/f_\pi$, $\theta_8$, $\theta_0$, lead to values of $\delta$ that are consistent with those obtained in the one-angle scheme (\ref{deta}). It is worth noting that similar estimates for the $\eta'\to\pi^+\pi^-\gamma$ decay (not shown in Table\,\ref{table1}) clearly indicate that the linear polynomial $P_{\eta'}(s_{\pi\pi})$ is insufficient for the two-angle scheme. In this case, we arrive at a consistent description only for large values of the $\Gamma_{\eta'\to\pi^+\pi^-\gamma} \simeq 70\,\mbox{keV}$.


 \section{$\eta^{(\prime)}\to\pi^+\pi^-l^+l^-$ decays}

Now let us move on to the main part of our calculations. The $\eta\to\pi^+\pi^-e^+e^-$ amplitude can be obtained from the $\eta\to\pi^+\pi^-\gamma$ decay amplitude (\ref{tot1}). For this, it is necessary to additionally take into account the decay of a photon into a lepton-antilepton pair. The corresponding vertex is well-known: $\mathcal L_{\gamma l l}=e A_\mu l^\mu$, where $l^\mu = \bar{l} \gamma^\mu l$ is the lepton current. As a result, we find
\begin{equation}
\label{amplitude1}
    \mathcal{M_\eta}  =  A_\eta \epsilon_{\mu\nu\alpha\beta}q^\nu p_+^\alpha p_-^\beta 
    \left[1  - \frac{3}{a} + \delta + \frac{3}{a}\,F_V(s_{\pi\pi})
    \left(\frac{m^2_\rho}{m_\rho^2 - q^2}\right) \right] \frac{e }{q^2}  \bar u(q_-)\gamma^\mu v(q_+)  \\
\end{equation}
where $p_\pm$ are the momenta of $\pi^\pm$, $s_{\pi\pi}= (p_{+}+p_{-})^2$, respectively $q_\pm$ are the leptons momenta and $q = q_{+}+q_{-}$ is momenta of virtual photon.

\begin{figure*}[t]
 \centering
  \begin{subfigure}{0.5\textwidth}
   \centering
   \begin{tikzpicture}
    \begin{feynman} 
      \vertex (a) {\(\eta^{(\prime)} \)};
      \vertex [dot, right=of a] (b) {};
      \vertex [dot, above right=of b] (c) {};
      \vertex [dot, below right=of b] (d) {};
      \vertex [dot, right=2.0cm of b] (e) {};
      \vertex [right=1.3cm of e] (f) {\(\pi^+ \)};
      \vertex [right=1.3cm of d] (g) {\(\pi^- \)};
      \vertex [dot, right=1.2cm of c] (n) {};
      \vertex [dot, right=1.2cm of n] (h) {};
      \vertex [above right=1.2cm of h] (i) {\(l^+\)};
      \vertex [below right=1.2cm of h] (j) {\(l^-\)};
      \diagram* {
        (a) -- [scalar] (b),
        (b) -- [fermion] (c),
        (c) -- [fermion] (e),
        (e) -- [fermion] (d),
        (d) -- [fermion] (b),  
        (n) -- [photon, edge label'=\(\gamma^* \)] (h),         
        (c) -- [double, edge label'=\(\quad \rho^0 \)] (n),
        (e) -- [scalar] (f),
        (d) -- [scalar] (g),
        (h) -- [anti fermion] (i),
        (h) -- [fermion] (j),
      };
     \end{feynman}
    \end{tikzpicture}
  \end{subfigure}%
 \centering
  \begin{subfigure}{0.5\textwidth}
   \centering
   \begin{tikzpicture}
    \begin{feynman} 
      \vertex (a) {\(\eta^{(\prime)} \)};
      \vertex [dot, right=1.2cm of a] (n) {};
      \vertex [dot, right=1.2cm of n] (b) {};
      \vertex [dot, above right=of b] (c) {};
      \vertex [dot, below right=of b] (d) {};
      \vertex [right=1.3cm of d] (e) {\( \partial \pi^{\pm} \)};
      \vertex [dot, below right=1.2cm of c] (g) {}; 
      \vertex [above right=1.2cm of c] (h) {\(\pi^{\mp} \)};
      \vertex [above right=1.2cm of g] (f) {\(l^+\)};
      \vertex [below right=1.2cm of g] (k) {\(l^-\)};
      \diagram* {
        (a) -- [scalar] (n),
        (n) -- [double, edge label'=\(f_1 \)] (b), 
        (b) -- [anti fermion] (c),
        (c) -- [anti fermion] (d),
        (d) -- [anti fermion] (b),  
        (d) -- [scalar] (e),
        (c) -- [scalar] (h),
        (c) -- [photon, edge label=  \( \gamma^* \)] (g), 
        (g) -- [anti fermion] (f),
        (g) -- [fermion] (k),
      };
     \end{feynman}
    \end{tikzpicture}
  \end{subfigure}%
 \caption{A box diagram describing the direct (non-resonant) production of final states in the NJL model with vector meson dominance. The box graph is affected by PA mixing effects (not shown). A triangle (not VMD-type) quark loop diagram is responsible for restoring low-energy theorems broken by PA transitions. Owing to the shift ambiguity related to the formal linear divergence of the one-loop triangle integral, the final result depends on the undetermined coupling $\delta$, which survives in the final expression (\ref{amplitude1}).}
 \label{diagram1}
\end{figure*}%

Let us make the necessary clarifications regarding the formula (\ref{amplitude1}). The origin of the first three terms in square brackets has already been discussed in detail above. The diagrams whose contact contributions lead to this result are shown in Fig.\,\ref{diagram1}. They represent the weight of the box $SU(3)\times SU(3)$ chiral anomaly, $(1-3/a+\delta)$, which in a special case (i.e., in the chiral limit $\delta=0$ and the KSFR relation $a=2$) transforms into Cohen's result $-1/2$ \cite{Cohen:1989}.
The fourth term is the contribution of the local part of triangle diagram shown in Fig.\,\ref{diagram2}. This term, at zero momenta, contributes with a weight of $3/a$. At non-zero momenta, the exchange of virtual $\rho^0$ mesons is described by the VMD vector form factor $F_V(s_{\pi\pi})$. In the case of the $\rho^0\to\gamma^*\to l^+ l^- $ transition, for kinematic reasons we neglect the imaginary part of $F_V(q^2)$ (actually we were convinced of this by conducting appropriate calculations both taking into account the finite width of the $\rho$-meson and when ignoring it).

\begin{figure*}[t]
\centering
  \begin{tikzpicture}
    \begin{feynman} 
      \vertex (a) {\(\eta^{(\prime)} \)};
      \vertex [dot, right=of a] (b) {};
      \vertex [dot, above right=of b] (c) {};
      \vertex [dot, below right=of b] (d) {};
      \vertex [dot, right=1.2cm of c] (n) {}; 
      \vertex [dot, right=1.2cm of n] (e) {};
      \vertex [dot, right=1.2cm of d] (f) {};
      \vertex [above right=1.3cm of f] (g) {\(\pi^+ \)};
      \vertex [below right=1.3cm of f] (h) {\(\pi^- \)};
      \vertex [above right=1.3cm of e] (i) {\(l^+\)};
      \vertex [below right=1.3cm of e] (j) {\(l^-\)};
      \diagram* {
        (a) -- [scalar] (b),
        (b) -- [fermion] (c),
        (c) -- [fermion] (d),
        (d) -- [fermion] (b),  
        (c) -- [double, edge label'=\(\rho^0 \)] (n), 
        (n) -- [photon, edge label'=\(\gamma^* \)] (e),   
        (d) -- [double, edge label'=\(\rho^0 \)] (f),
        (f)  -- [scalar] (g),
        (h) -- [scalar] (f),
        (e) -- [anti fermion] (i),
        (e) -- [fermion] (j),
      };
     \end{feynman}
    \end{tikzpicture}
\caption{Triangle quark-loop diagram in the NJL model with vector meson dominance contributing to $\eta^{(\prime)} \to\pi^+ \pi^- l^+l^-$ decays. This diagram is not affected by the PA mixing which is forbidden due to the Landau-Yang theorem.}
 \label{diagram2}
\end{figure*}

Expression (\ref{amplitude1}) in the particular case of $\delta=0$ and $a = 2$ coincides with the amplitude obtained in \cite{Picciotto:93} within the framework of the HLS approach. These two seemingly (but only seemingly) insignificant differences essentially affect the final result, as they lead to noticeable changes in the structural (polynomial) part of the $\eta^{(\prime )}\to\pi^+\pi^-\gamma$ amplitudes, namely, they more than double the values of the slope parameters $\alpha^{(\prime )}$. Our task now is -- to explore how these changes affect the $\eta^{(\prime)}\to\pi^+\pi^- l^+ l^-$ decay widths. 

Using the amplitude (\ref{amplitude1}) one can find the differential decay rate of $\eta^{(\prime)}\to\pi^+\pi^- l^+ l^-$ in terms of the variables $s_{\pi\pi}$ and $q^2$
\begin{equation}
\label{diff}
d\Gamma =  \frac{|M_P|^2 }{m_P^3 3^2 2^{14}\pi^5}\sigma_\pi^3\sigma_l(3-\sigma_l^2)\lambda^{\frac{3}{2}}(m^2_P, s_{\pi\pi}, q^2) s_{\pi\pi} ds_{\pi\pi}\frac{dq^2}{q^2},
\end{equation}
where $m_P$ is a mass of the pseudoscalar state $P=\eta, \eta'$, and $m_l$ is a mass of lepton $l=e,\mu$, other notations are
\begin{eqnarray}
&&\lambda(a,b,c)=a^2+b^2+c^2-2(ab+ac+bc), \nonumber \\
&&\sigma_\pi=\sqrt{1-\frac{4m_\pi^2}{s_{\pi\pi}}}, \quad  \sigma_l =\sqrt{1-\frac{4m_l^2}{q^2}}, \nonumber \\
&&M_P= eA_P\left[1  - \frac{3}{a} + \delta + \frac{3}{a}\,F_V(s_{\pi\pi}) \left(\frac{m^2_\rho}{m_\rho^2 - q^2}\right) \right].
\end{eqnarray}

For numerical estimates we use the values of $\delta^{(\prime)}$ given in (\ref{deta}) and (\ref{detap}). The obtained branching ratios for different $\eta^{(\prime)}\to\pi^+\pi^- l^+ l^-$ channels, as well as a comparison of the obtained results with experimental data \cite{PDG:24} and calculations within the framework of alternative approaches that take into account coupled meson-meson channels \cite{Borasoy:2007}, or using ChPT and standard unitarization methods \cite{Kubis:22}, are given in Table\,\ref{table2}.

Our calculations show that the results are generally consistent with the data enshrined in the particle data tables. The small ($1.6\sigma$) discrepancy observed in mode $\eta'\to\pi^+\pi^- e^+ e^-$ is statistically insignificant. We also note that the experimentally established upper limit for $\eta\to\pi^+\pi^- \mu^+ \mu^-$ decay has now shifted to Br($\eta\to\pi^+\pi^- \mu^+ \mu^-)<4.0\times10^{-7}$ \cite{BESIII:2025cky}, which turned out to be three orders of magnitude lower than the averaged data presented by PDG. Theoretical estimates suggest that the upper bound may be even lower.

\begin{table}[tbp!]
\caption{The branching ratios of anomalous decays $\eta^{(\prime)}\to\pi^+\pi^-l^+l^-$ obtained in the model under consideration with a one-angle mixing scheme for $\eta$-$\eta'$ mixing}
\begin{center}
\small
\begin{tabular}{lcccc}
\hline
\hline
Decay mode
& Model 
& PDG  \cite{PDG:24} 
& \cite{Borasoy:2007}
& \cite{Kubis:22}
\\
\hline
$\eta\to\pi^+\pi^-e^+e^-$     
& $2.69(15)\times10^{-4}$ 
& $2.68(11)\times10^{-4}$
& $2.99^{+0.06}_{-0.09}\times 10^{-4}$ 
& $2.65(17)\times10^{-4}$ 
\\
$\eta\to\pi^+\pi^-\mu^+\mu^-$ 
& $7.26(46)\times10^{-9}$ 
& $<3.6\times10^{-4}$ 
& $7.5^{+1.8}_{-0.7}\times 10^{-9}$
& $6.36(39)\times10^{-9}$ 
\\
$\eta'\to\pi^+\pi^-e^+e^-$     
& $2.20(10)\times10^{-3}$ 
& $2.42(10)\times10^{-3}$ 
& $2.13^{+0.17}_{-0.31}\times 10^{-3}$
& $2.21(15)\times10^{-3}$ 
\\
$\eta'\to\pi^+\pi^-\mu^+\mu^-$ 
& $1.90(26)\times10^{-5}$ 
& $1.9(4)\times10^{-5}$ 
& $1.57^{+0.40}_{-0.47}\times 10^{-5}$
& $2.25(14)\times10^{-5}$
\\
\hline
\hline
\end{tabular}
\label{table2}
\end{center}
\end{table}

\begin{figure}
\includegraphics[width=1.00\textwidth]{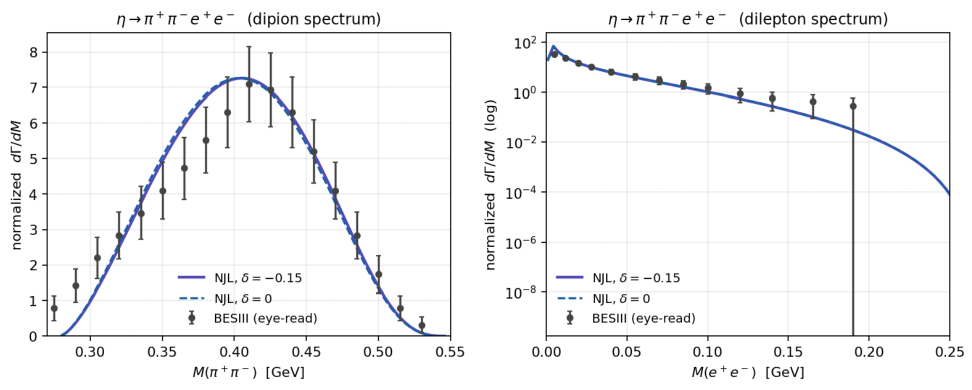}
\caption{Model prediction for the invariant mass distribution of the $\pi^+\pi^-$ (left panel) and $e^+e^-$ (right panel) pairs for the $\eta\to\pi^+\pi^- e^+ e^-$ decay mode with $\delta = -0.15$ (solid) and $\delta = 0$ (dashed). Both curves are normalised to unit area. Overlaid points are visual estimates of the BESIII histograms in Fig. 5 of \cite{BESIII:2025cky}, similarly normalised; error bars are $\sqrt N$ statistical only.}
\label{f4}      
\end{figure}

The individual distributions can be computed by integrating (\ref{diff}) with respect to $q^2$ or $s_{\pi\pi}$, and the results are shown in Fig.\,\ref{f4}. These are what BESIII, KLOE-2, and JEF actually measure. The invariant mass distribution $M(\pi^+\pi^-)$  (left panel) shows a broad bump peaking near $0.40\,\mbox{GeV}$. The shape is set by $\rho$-meson dominance via $F_\rho (s_{\pi\pi})$ (see Eq.\,(\ref{width})) cut off by $\eta$ phase space --  the nominal $\rho$ pole at $0.775\,\mbox{GeV}$ is outside the kinematic boundary ($M_{\pi^+\pi^-}\leq m_\eta \simeq 0.548\,\mbox{GeV}$), so we see only the rising edge of the resonance modulated by $\sigma^3_\pi\lambda^{3/2}$. The $\delta = -0.15$ and $\delta=0$ curves are nearly indistinguishable in shape -- the surface term moves the normalisation by $\sim 36\%$, not the spectrum. The invariant mass distribution $M(e^+ e^- )$ (right panel) is plotted on a logarithmic scale: steep fall from threshold, dominated by the $1/q^2$ photon propagator. Every dilepton conversion process inherits this characteristic shape regardless of hadronic dynamics. It is clear from the plots that the differential distributions are insensitive to the value of the parameter $\delta$, which is why the values of the $\delta^{(\prime)}$ were fixed based on the integral data -- the $\eta^{(\prime)}\to \pi^+\pi^-\gamma$ decay widths.


\section{Conclusions}

In this paper, we extended the study of anomalous $\eta^{(\prime )}\to\pi^+\pi^-\gamma$ decays initiated in \cite{Osipov:2020vad} to semileptonic decays $\eta^{(\prime)}\to\pi^+\pi^- l^- l^+$. In summary, we will highlight the main results obtained.  

1. We find interesting the relationship between the parameters $\delta^{(\prime )}$ and $\alpha^{(\prime )}$, expressed by the formula (\ref{alpha2}). This relation is a consequence of using the VDM-form to describe the pion's vector form factor.
This new relation implies that the slope parameter $\alpha^{(\prime)}$ appears to contain important information about the explicit chiral symmetry breaking at an anomalous VAAA-type vertex. Moreover, this distorts the balance between the contributions of box and triangle anomalies to the amplitude $\eta^{(\prime)}\to\pi^+\pi^-\gamma$ in favor of the box one giving
$$
A^{tot}=\frac{3}{2}A^{VVA}+\left(-\frac{1}{2}+\delta^{(\prime )}\right)A^{VAAA}. 
$$
In chiral approaches with vector mesons (massive Yang-Mills, hidden local symmetry), the parameter $\delta^{(\prime )}=0$. Therefore, its appearance in the NJL model may seem unexpected at first glance. However, this is not surprising. Based on the fact that in the NJL model the effective vertices of mesons are represented by the local part of one-loop quark diagrams, we showed that the parameter $\delta^{(\prime )}$ directly owes its origin to the contribution of a triangle quark diagram, which on the one hand leads to a deviation from the VMD scheme, and on the other belongs to the class of finite, but formally linearly divergent diagrams, the shift of the integration variable in which gives rise to the appearance of a surface term. It is this circumstance that underlies the non-zero value of parameter $\delta^{(\prime )}$.     

2. The established link between the parameters $\delta^{(\prime )}$ and $\alpha^{(\prime)}$ allows us to find the value of the parameter $\alpha^{(\prime)}$ not from spectral data, as is usually done, but from integral data -- the partial widths of the corresponding radiative decays that determine $\delta^{(\prime )}$.

3. Based on the known values of the $\eta$-$\eta'$ mixing parameters and the phenomenological values of the $\eta^{(\prime )}\to\pi^+\pi^-\gamma$ decay widths, the values of $\delta^{(\prime )}$ (and hence $\alpha^{(\prime )}$) were determined and compared with the results of alternative $\alpha^{(\prime )}$ estimates obtained from the analysis of spectral data. A comparison showed that the slope parameter values calculated using formula (\ref{alpha2}) are in complete agreement with the spectral measurement data. As an alternative, a two-angle mixing scheme was also considered. There is also good agreement with alternative approaches here, but in the case of $\eta'\to\pi^+\pi^-\gamma$ decay this agreement is achieved only at higher values of the decay width. Final conclusions here require additional theoretical analysis (for example, increasing the polynomial part of the amplitude to its quadratic form), which is possible, but we have postponed this issue for the future.

4. While data on one-photon $\eta^{(\prime )}\to\pi^+\pi^-\gamma$ decays allow us to completely record the values of the low-energy constants of the model, semileptonic $\eta^{(\prime )}\to\pi^+\pi^- l^- l^+$ decays allow us to check how true our ideas about the structural part of the hadronic amplitude are at non-zero photon virtualities. The model prediction reproduces the qualitative shape correctly: $\rho$-dominated $\pi^+\pi^-$ peaking at the high end of phase space and $1/q^2$ fall in the $e^+ e^-$ spectrum. The eye-read BESIII points sit on top of the predicted curves within their (large) approximate uncertainties. Critically, the shape does not discriminate between $\delta=-0.15$ and $\delta=0$ -- only the integrated rate does. That is exactly why we have to anchor $\delta$ from the $\eta^{(\prime)}\to\pi^+\pi^-\gamma$ partial width rather than from the dilepton spectrum itself.

\subsection*{Acknowledgements}
This research has been funded by the Science Committee of the Ministry of Science and Higher Education of the Republic of Kazakhstan Grant No. AP32318769.


\end{document}